\documentclass[10pt]{iopart}
\usepackage{iopams}
\usepackage{graphicx}
\usepackage{bm}
\usepackage{mathrsfs}
\usepackage{iopams,setstack}

\newcommand{\beq}{\begin{equation}}
\newcommand{\eeq}{\end{equation}}
\newcommand{\beqa}{\begin{eqnarray}}
\newcommand{\eeqa}{\end{eqnarray}}
\def\nn{\nonumber\\}
\def\eq#1{(\ref{#1})}

\def\st{space-time}

\def\Sch{Schwarzschild}

\def\text#1{\rm #1}

\begin{document}

\title{The white wall, a gravitational mirror
}

\author{Ricardo\ E.  Gamboa Sarav\'{\i}\dag\  \ddag
}

\address{\dag\ Departamento de F\'{\i}sica, Facultad de Ciencias
Exactas, Universidad Nacional de La Plata, Argentina. }
\address{\ddag\ IFLP, CONICET,
Argentina.  }

\ead{quique@fisica.unlp.edu.ar}

\date{\today}

\begin{abstract}
We describe the  exact solution of  Einstein's equation
corresponding to  a static homogenous distribution of matter with
plane symmetry lying below $z=0$. We study the geodesics in it and
we show that this simple spacetime exhibits very curious
properties. In particular, it has a repelling singular boundary
and all geodesics bounce off it.
\end{abstract}



\section{Introduction}

Given  the complexity of Einstein's field equations, one cannot
find exact solutions except in spaces of rather high symmetry, but
very often with no direct physical application. Nevertheless,
exact solutions can give an idea of the qualitative features that
could arise in General Relativity  and so, of possible properties
of realistic solutions of the field equations.

In this paper we want to illustrate some curious features of gravitation by means of a very simple  solution:  the 
gravitational field  of a static homogenous distribution of matter
with plane symmetry lying below $z=0$.

 Of course, the Newtonian exterior (i.e., $z>0$) solution $\bm{g}=-g\, \bm{e}_z$ is very simple:  a static {\em uniform}
gravitational field pointing in the negative $z$ direction.

This  uniform gravitational field  has played  an outstanding role
in the History of Physics and of Mankind, being perhaps the most
important physical entity in our daily life, since it is the field
that we always  feel and to which evolution has made us perfectly
adapted.

It is not by chance that Galileo started his studies  of movement
laws by analyzing the way  bodies move in its presence, or that
Einstein did explicit use of it in his crucial Equivalence
Principle in 1911, the cornerstone of his wonderful travel from
flat to curved \st.

In spite of this, the exact solution of Einstein's field equations
representing it, {\em i.e.}, a full general relativistic (plane)
homogenous gravitational field in the vacuum,  seems not to have
been properly spread into the wide physics community. In fact, we
haven't found it mentioned in any book on General Relativity.
However,  this simple vacuum solution has been known for several
years. It appears in Taub's study of Mach's Principle in General
Relativity \cite{taub1}, relativistic perfect fluids with plane
symmetry \cite{taub}, or in the study of domain walls in the early
Universe \cite{vil}.

This Taub's plane vacuum  solution turns out to be the exterior
gravitational field (i.e., $z>0$)
of the problem we are considering here: the  solution of
Einstein's equation
corresponding to  a static homogenous distribution of matter with plane symmetry lying below $z=0$. 
For these reasons and for the sake of completeness,
we present the solution in Sec. II, and a detailed study of the
geodesics  in it in Sec. III.

This very simple spacetime turns out to present some  somehow
astonishing properties. Namely, the attraction of the {\em
infinite} amount of matter lying below $z=0$ shrinks the \st\ in
such a way that it finishes high above at a singular boundary.

Throughout this paper, we adopt the convention in which the \st\
metric has signature $(-\ +\ +\ +)$,  the  system of units in
which the speed of light $c=1$, 
and $g$ will denote gravitational field and not the determinant of
the metric.

\section{The metric}

 We start by determining the solution  of Einstein's
equation corresponding to the exterior gravitational field of a
static and homogenous distribution of matter with plane symmetry
sitting somewhere at $z<0$. That is, it should be invariant under
translations in the plane and under rotations around its  normal.

  Since the solution should be static, we can find a time
coordinate $t$ such that all metric's components are
$t$-independent, and in addition $g_{ti}=0$ for $i=1,2,3$. By the
plane translation symmetry, the metric can depend only on the
variable $z$, so \beq ds^2= -A(z)\,dt^2 + h_{ij}(z)\,dx^i\
dx^j,\eeq where $i,j=1,2,3$ and $x^3=z$. Furthermore, the rotation
invariance imposes $h_{11}=h_{22}$ and $h_{12}=h_{21}=0$.
Moreover, by a  change of the plane coordinates of the form
$x\rightarrow x+f(z)$, and $y\rightarrow y+g(z)$ we can make
$h_{13}=h_{23}=0$. So, we can choose coordinates $x, y$ on the
plane in such a way that $g_{xy}=g_{xz}=g_{yz}=0$, and
$g_{xx}=g_{yy}$. Thus, we have found coordinates $(t,x,y,z)$ such
that \beq ds^2= -A(z)\,dt^2 + B(z)\,(dx^2+dy^2)+ dz^2,\eeq where,
without loss of generality, we have chosen $g_{zz}=1$, so that the
coordinate $z$ turns out to be the physical distance to the plane
$z=0$.

The  non identically zero components of  the Einstein tensor are
\beqa G_{tt}=\frac{A(z)\,\left( {B'(z)}^2 - 4\,B(z)\,B''(z)
\right) }{4\,{B(z)}^2},\\ \nn G_{xx}=G_{yy}=-\frac{
{A(z)}^2\,{B'(z)}^2 + {B(z)}^2\, {A'(z)}^2 }{4\,{A(z)}^2\,B(z)}+
\nn \,\frac{2\,B(z)\,A''(z)+ A'(z)\,B'(z) + 2\,A(z)\,B''(z)
 }{4\,{A(z)}},\\ \nn G_{zz}=\frac{B'(z)\,\left(
2\,B(z)\,A'(z) + A(z)\,B'(z) \right) }{4\,A(z)\,{B(z)}^2}.\eeqa

The vacuum Einstein equation, i.e.,  $G_{ab}=0$, can be readily
solved for $z>0$. In fact, as the metric depends only on one
variable $z$, it corresponds to a Kasner solution with exponents
$\{-1/3,2/3,2/3\}$ (see, for example, \cite{land}).

The general solution is \beqa A(z)=C_1\,{\left(C_3 + z \right)
}^{-\frac{2}{3}},\, \,\,\,\,\text{and} \,\, \,\,
B(z)=C_2\,{\left(C_3+ z \right) }^{\frac{4}{3}} \eeqa for
arbitrary constants $C_1$, $C_2$ and $C_3$. Now, we fix the
constants by demanding 
in a neighborhood of the plane $z=0$ \beq
g_{tt}\simeq-(1+2\Phi(z))=-(1+2g z)\,\,\,\,\text{for}\,\,\,\,|g
z|\ll 1.\eeq Thus, for $z>0$ the metric reads \beq \label{met}
ds^2= -\frac{1}{(1-3g z)^{2/3}}\,dt^2 + (1-3g z)^{4/3}(dx^2+dy^2)+
dz^2,\eeq this is the Taub's plane vacuum solution \cite{taub1}.
Notice that the coordinates change $\hat{z}=(1-(1-3gz)^{4/3})/4g$
brings the metric (\ref{met}) to the  form  used in references
\cite{taub1,taub,vil}\beq\label{taucoord} ds^2= \frac{(-dt^2+
d\hat{z}^2)}{{(1- 4 g \hat{z})^{1/2}}} + (1- 4 g
\hat{z})(dx^2+dy^2).\eeq

By matching the solution of Einstein's equation for the interior
of the matter (i.e., $z<0$) with the exterior one Eq.(\ref{met}),
we can compute the constant $g$ from the matter properties
\cite{taub}. But, apart from the above comment, matter underground
is not playing any relevant role here, because the outside
solution will always be the  one given in Eq.(\ref{met}).

 Thus, this metric is the unique   exact solution of
vacuum Einstein's equations satisfying the required plane symmetry
up to a coordinate transformation. As it occurs with
Schwarzschild's case, it depends only on a length scale parameter
$1/g$. The metric (\ref{met}) becomes the Minkowskian one when
$g=0$. Otherwise, it represents a curved spacetime and it turns
out to have a \st\ curvature singularity on the ``plane''
$z={1/3g}$,  since straightforward computation of the scalar
quadratic in the Riemann tensor, yields
 \beq \label{RR}
R_{abcd}R^{abcd}=\frac{192\,g^4}{{\left( 1 - 3gz \right) }^4},
\eeq where $a,b,c,d=0,1,2,3$.

Note that each spacelike slice of \st \  $t=t_0$ and $z=z_0$ is a
Euclidean plane with metrics \beq  d\ell^2=(1-3g
z)^{4/3}(dx^2+dy^2),\eeq so its ``size''  contracts when $z$
increases  and becomes a point at the singularity ($z=1/3g$).

Notice that, beyond the singularity (i.e., $z>1/3g$), a mirror
copy of the empty \st\  emerges. But, as we shall show below, no
geodesic can  go from one to the other.

 In the next section we shall study the geodesics in this space time.
 For the sake of completeness, we shall consider the complete Taub plane spacetime, i.e., $-\infty<z<1/3g$, disregarding  the matter lying below $z=0$. Of course, it should be understood than when a geodesic reaches the matter it  is modified in  some way, but here we are only interested in what happens in the vacuum ($0<z<1/3g$).

Notice that, the coordinates change $t'=(1-3gz)^{-1/3}t$,
$x'=(1-3gz)^{2/3}x$, $y'=(1-3gz)^{2/3}y$, and $z'=z$ brings the
vacuum  metric (\ref{met}) to the  form \beqa \fl\label{iso}ds^2=
-dt'^2 +dx'^2+dy'^2+ dz'^2 \nn +g\ \frac{2 t' dt' dz' +4 x' dx'
dz'+4 y' dy' dz'}{(1-3g z)}+ 
g^2\ \frac{4 x'^2+4
y'^2-t'^2}{(1-3g z)^{2}}\ dz'^2,\eeqa which shows that Taub's
plane symmetric  spacetime is asymptotically flat when
$z\rightarrow-\infty$. Of course, we see from (\ref{RR}) that it
is not flat for finite values of $z$.

\section{The Geodesics}

We want to study the geodesics in this \st . Since the metric is
independent of $t$, $x$ and $y$, the momentum covector components
$p_t$, $p_x$ and $p_y$ are constant along the geodesics. For
timelike geodesics, we choose $\tau$ to be the proper time while
for null ones, we choose $\tau$ to be an affine parameter. So, we
can write \beqa  \label{dz}\left(\frac{dz}{d\tau}\right)^2={\left(
1 - 3gz \right) }^{{2/3}} {\tilde{E}}^2 -\mu -
\frac{(\tilde{p_x}^2+\tilde{p_y}^2)}{{\left( 1 - 3gz \right)
}^{4/3}},\label{rebote}\\
 \label{dt}\frac{dt}{d\tau}={\left( 1 - 3gz \right) }^{{2/3}}
{\tilde{E}},\\\label{dx}\frac{dx}{d\tau}={\left( 1 - 3gz \right)
}^{{-4/3}} {\tilde{p}_x},\\\frac{dy}{d\tau}={\left( 1 - 3gz
\right) }^{-4/3} {\tilde{p}_y}, \eeqa where $\mu=1$,
${\tilde{E}}=-p_t/m$, $\tilde{p_x}=p_x/m$ and $\tilde{p_y}=p_y/m$
for timelike geodesics; and $\mu=0$, ${\tilde{E}}=-p_t$,
$\tilde{p_x}=p_x$ and $\tilde{p_y}=p_y$ for null  ones.

The right hand side of (\ref{rebote}) must be positive, so we see
that there is a bouncing point for each geodesic. To get the $z$
coordinate of such point, we solve $({dz}/{d\tau})^2=0$ in
Eq.(\ref{rebote}). In all cases, this yields  a cubic equation
with only one real root .

\subsection{Null Geodesics}

 It follows from (\ref{rebote}) that the maximal height that a null geodesic reaches is \beq \label{h}h_{max}=
\frac{1}{3g}\left(1-{\tilde{p}_h}/{\tilde E}\right)\,, \eeq  where
$\tilde {p}_h=\sqrt{\tilde p_x^2+\tilde p_y^2}$. Thus,
nonvertical null geodesics  bounce before getting to the
singularity, whereas vertical null ones just touch  it.

In this case, the geodesics equation can be integrated in a closed
form. Indeed, from (\ref{dz}), (\ref{dt}) and (\ref{dx}) we obtain
the equations governing these geodesics \beqa \frac{dt}{dz}=\pm
\frac{\left( 1 - 3gz \right)^{{4}/{3}}}{\sqrt{ {\left( 1 - 3gz
\right) }^2-(\tilde{p}_h/\tilde{E})^2}}\,,\\ \frac{dx}{dz}=\pm
\frac{(\tilde{p}_h/\tilde{E})\left( 1 - 3gz
\right)^{-{2}/{3}}}{\sqrt{ {\left( 1 - 3gz \right)
}^2-(\tilde{p}_h/\tilde{E})^2}}\,,\\ \frac{dy}{dz}=0 ,\eeqa where,
without loss of generality, we have assumed $\tilde p_y=0$. The
solution of these equations can be written as \beqa\fl\label{t}
|t- t_0| = (\tilde{p}_h/\tilde{E})^{4/3}\
\frac{\sqrt{\pi}\;\Gamma(-2/3)}{6g\,\Gamma(-1/6)} +\frac{( 1 - 3gz
)^{{4}/{3}}}{4g}\,\,
_2F_1\!\left(-\frac{2}{3},\frac{1}{2};\frac{1}{3};\frac{(\tilde{p}_h/\tilde{E})^2}{(
1 - 3gz )^2}  \right)\,,\\ \fl\label{x} |x- x_0| =
(\tilde{p}_h/\tilde{E})^{1/3}\
\frac{\sqrt{\pi}\;\Gamma(1/3)}{6g\,\Gamma(5/6)}-\frac{(\tilde{p}_h/\tilde{E})}{2g
( 1 - 3gz )^{{2}/{3}}}\,\,
_2F_1\!\left(\frac{1}{3},\frac{1}{2};\frac{4}{3};\frac{(\tilde{p}_h/\tilde{E})^2}{(
1 - 3gz )^2} \right)\,,\\ \fl y=y_0 , \eeqa where $_2F_1(a,b;c;z)$
is the Gauss hypergeometric function (see for example
\cite{abra}).

From \eq{t} and \eq{h} we readily see that all non-vertical null
geodesics start and finish at $z=-\infty$, and all of them reach a
turning point at a finite distance from the singularity, and the
smaller their horizontal momentum is, the closer they get near the
singularity.

Now, let us consider two points on the plane $z=0$ connected by a
null geodesic, we readily get from \eq{x} that the distance
between them is  \beq (\tilde{p}_h/\tilde{E})^{1/3}\
\frac{\sqrt{\pi}\;\Gamma(1/3)}{3g\,\Gamma(5/6)}-\frac{(\tilde{p}_h/\tilde{E})}{g
}\,\,_2F_1\!\left(\frac{1}{3},\frac{1}{2};\frac{4}{3};(\tilde{p}_h/\tilde{E})^2
\right)\,. \eeq This  distance vanishes for
$\tilde{p}_h/\tilde{E}=0$ and for $\tilde{p}_h/\tilde{E}=1$,
furthermore we can easily compute that, for
$\tilde{p}_h/\tilde{E}=0.302684\dots$, it reaches a maximum \beq
R= \frac{0.635164\dots}{g}\,.\eeq Therefore an observer standing
on the plane can only see the points of the plane contained in a
circle of radius $R$. Of  course, a similar result holds for any
plane  $z= constant.$

In Fig.1 we show the null geodesics that can reach a point $O$.
The region of the space that can be seen by an observer at rest at
the point  $O$ is bounded by a ``cupola'', and he cannot see
anything that is beyond it. Moreover, any point $P$ in this region
can be seen from $O$ by looking at two different directions. But,
since both rays of light last different   times to travel from $P$
to $O$, the observer sees what happens at $P$ at two different
times. For instance, the observer can see his neighborhood ``at
present'' by looking horizontally, whereas he see what happened
there ``$1/2g$  ago'' by looking vertically.

\begin{figure}[h]
\begin{center}
\includegraphics[width=\textwidth]{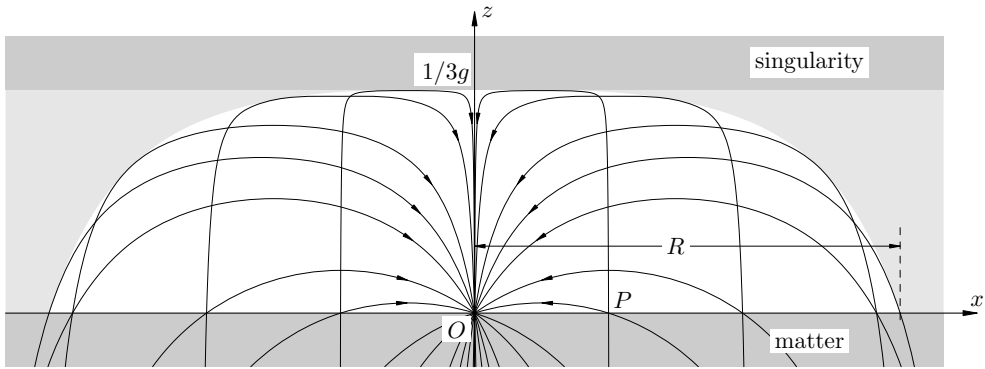}
\caption{\label{fig }Null geodesics that reach the the point $O$.
}\end{center}
\end{figure}

When $\tilde p\rightarrow0$ we can get from \eq{t} and \eq{x} the
limit geodesic for vertically moving photons

\beq\fl\label{phot} |t- t_0| = \frac{1}{4g }{\left( 1 - 3gz
\right) }^{\frac{4}{3}}\,,\hspace{2cm}
x=x_0\hspace{1cm}\text{and}\hspace{1cm}y=y_0\ . \eeq Thus, we see
that a photon moving upwards crossing the plane $z=0$ at
coordinate time $t=t_0-1/4g$ reaches the singularity at time
$t_0$,  comes back to $z=0$ at $t_0+1/4g$, and  afterwards it
follows its travel to $z=-\infty$. Thus, every null geodesic
traveling upwards is
 reflected at the singularity.

Therefore, we can fill the whole \st\ with  never-stopping
future-oriented  null geodesics, all of them starting  and
finishing at $z=-\infty$.

Now, consider a local Galilean observer  momentarily at rest at a
height $z$  measuring the energy of a photon with
$p_t=-\tilde{E}$. Since the time component of the observer's
velocity must be $U_{\!obs.}^{\ \ \ t} =( 1 - 3\,g\,z)^{1/3}$, for
$g_{ab}\ U_{\!obs.}^{\ \ \ a}\ U_{\!obs.}^{\ \ \ b}=-1$, the
energy he measures is $-U_{\!obs.}^{\ \ \ a}\ p_a=-U_{\!obs.}^{\ \
\ t}\ p_t= (1 - 3\,g\,z)^{1/3}\ \tilde{E}$. Since $g_{tt}=-1$ at
$z=0$, the constant of motion $p_t=-\tilde{E}$ is the energy that
an observer at $z=0$ would measure.  So, if a photon is emitted at
height $h$ and received at $z=0$, we have \beq\label{nu} \nu_{rec}
= \frac{ \nu_{em}}{(1-3g h)^{1/3}},\eeq and if the photon were
emitted at the singularity, the frequency  record would be
infinitely shifted to the blue.

On the other hand, for ``terrestrial" (i.e., $gz\ll 1$)
experiments, the exact result in Eq.(\ref{nu}) becomes the
classical Einstein prediction \beq \nu_{rec} \simeq {
\nu_{em}}{(1+g h)},\eeq measured by Pound, Rebka and Snider
\cite{prs}.

\subsection{Timelike Geodesics}

Clearly, \eq{dz} shows that no massive particle can reach the
singularity. For timelike geodesics, the general expression for
the maximal height reached can be explicitly written  down, but it
turns out to be rather involved. We show it here only for the sake
of completeness \beqa\fl \label{hmax}h_{max}= \frac{1}{3g}-
\frac{1}{9\sqrt{3}\,g\,{\tilde{E}}^3}{\left(1 +
\frac{1+\sqrt[3]{{\left( 1 + 27\,{\tilde{E}}^4\,{\tilde p}^2/2 +
3\,{\tilde{E}}^2\,{\sqrt{3{\tilde p}^2 +
81\,{\tilde{E}}^4\,{{\tilde p}}^4/4}} \right)^2 } }} {\sqrt[3]{{ 1
+ 27\,{\tilde{E}}^4\,{\tilde p}^2/2 +
3\,{\tilde{E}}^2\,{\sqrt{3{\tilde p}^2 +
81\,{\tilde{E}}^4\,{{\tilde p}}^4/4}} }}{}}\right)^{3/2}}.\nn
\eeqa So, for the sake of clarity, we shall discus two simple
cases:

 For particles moving
vertically (i.e. $\tilde{p_x}=\tilde{p_y}=0$), equation
(\ref{hmax}) reduces to \beq h_{max}=
\frac{1}{3g}\left(1-\frac{1}{\tilde E^3}\right). \eeq In this
case, from (\ref{dz}) and (\ref{dt}) we obtain the equation
governing these geodesics \beq \frac{dt}{dz}=\pm \frac{\left( 1 -
3gz \right)^{{2}/{3}}}{\sqrt{ {\left( 1 - 3gz \right)
}^{{2}/{3}}-{1}/{{\tilde{E}}^2}}}\ .\eeq The solution of this
equation can be written as
 \beqa \label{part}\fl  |t- t_0| = \frac{1}{4g \tilde{E}^4}{\,
 \left[{\left( 1 - 3gz \right) }^{{1}/{3}}{\tilde{E}}\,{\sqrt{ {\left( 1 - 3gz \right) }^{{2}/{3}}{\tilde{E}}^2-1}}\ \ \,{\left( {\left( 1 - 3gz \right) }^{{2}/{3}} {\tilde{E}}^2 +\frac{3}{2}\right) } \right.
} \nn+ \left.\frac{3}{2}\ {\log\left({\left( 1 - 3gz \right)
}^{{1}/{3}}{\tilde{E}}\, + {\sqrt{ {\left( 1 - 3gz \right)
}^{{2}/{3}}{\tilde{E}}^2-1}}\ \right)}\right]\ .\eeqa  Thus, since
$|t|\rightarrow\infty$ when $z\rightarrow-\infty$, we see that
vertical timelike geodesics start at $z=-\infty$,  reach the
highest point $z=h_{max}$ at coordinate time $t_0$, and  after
that they follow their return travel to $z=-\infty$. Notice that
(\ref{part}) readily shows how timelike geodesics   tend to null
ones (\ref{phot}) for  high-energy particles (i.e., $\tilde{E}\gg
1)$.

If we  call $v(z)$ the (tri)velocity of the particle  measured by
a local inertial observer  momentarily at rest at a height $z$, we
have \beq U_{\!obs.}^{\ \ \ a}\ U_a=g_{tt}\ U_{\!obs.}^{\ \ \ t}\
\frac{dt}{d\tau}=-{\left( 1 - 3gz \right)
}^{{1}/{3}}{\tilde{E}}=\frac{-1}{\sqrt{1-v(z)^2}}\ ,\eeq where
$U_{\!obs.}^{\ \ \ a}$ and $U^{a}=\frac{dx^a}{d\tau}$ are the
velocities of the observer and the particle repectively. So, it
holds \beq v(z)=\pm \sqrt{1-\frac{1}{\left( 1 - 3gz
\right)^{{2}/{3}}{\tilde{E}}^2}}\ , \eeq which shows that
$|v(z)|\rightarrow 1 $ when $z\rightarrow -\infty$.

For nonrelativistic particles, (i.e.,
$\tilde{p_x}^2+\tilde{p_y}^2<\tilde{p_x}^2+\tilde{p_y}^2+\tilde{p_z}^2\ll
1$),  equation (\ref{hmax}) reduces to  \beq h_{max}=
\frac{1}{3g}\left(1-\frac{1}{\tilde
E^3}-\frac{3}{2}{\tilde{E}(\tilde{p_x}^2+\tilde{p_y}^2)+\
O(\tilde{p})^4}\right). \eeq Now, by calling $v_x,v_y,v_z$  the
velocity components  at the plane $z=0$, we can write
$\tilde{E}\simeq1+(v_x^2+v_y^2+v_z^2)/2 $, $\tilde{p_x}\simeq v_x
$ and $\tilde{p_y}\simeq v_y $, getting the Galilean result \beq
h_{max}\simeq \frac {v_z^2}{2g}.\eeq

As all geodesics bounce, we see that the  singularity at $z=1/3g$
acts as a {\it gravitational mirror}, so we call it a {\em white
wall} and we  are able to  think of it by removing the point
$z=1/3g$ as the boundary of the \st.

Hence, the attraction of the {\em infinite} amount of matter lying
below $z=0$, shrinks the \st\ in such a way that it finishes high
above at the singular boundary.

Notice that, from the Newtonian point of view, we can think of a
plane as a sphere with infinite radius. Here, in contrast, we
cannot obtain the exact solution as a limit of \Sch's metric,
since globally these \st s are completely different.

\section{Concluding remarks}

We have described  the  general external solution of Einstein's
equation  produced by a static and homogenous distribution of
matter with plane symmetry sitting somewhere at $z<0$. The
solution turns out to have a repelling boundary high above and all
geodesics bounce off  it. This singularity is not the source of
the field, but it arises owing to the attraction of distant
matter.

Of course, we don't think there is place in our universe for an
object with such a symmetry, but this simple example  shows some
peculiar properties that perhaps could help us understand relevant
facts in Nature.

Regarding the interior solutions, since as long as  they exist,
the exterior solution inexorably is the Taub's one, they don't
play a relevant role in our analysis. However, as already
mentioned, given an equation of state  $\rho=(p)$, by matching
both solutions (Taub's plane and the interior one, e.g., at $z=0$)
$g$ can be computed from the matter properties \cite{taub}. For
instance, it can be shown that, for  matter with constant density
$\rho_0>0$, it can be done, and the space-time also finishes down
below at another singularity at  a finite depth \cite{gs}.

After we finished this work  we learned about reference
\cite{bed}, where some of the  results here presented can be
found. However, these authors consider a very different
space-time: the mirror-symmetric matching of two asymptotic flat
Taub's domains to the ``left" and ``right" of a planar shell with
infinite negative mass density sitting at the singularity, whereas
we are arguing here that Taub's plane solution is also the
external gravitational field of ordinary ($\rho$ and $p\geq0)$
matter sitting at negative values of $z$, and the singularity that
arises high above is due to the attraction of the distant matter.

\section*{Acknowledgements}
I would like to thank  Mariel Santangelo and  H\'ector Vucetich
for their helpful comments on this
 manuscript.

\section*{References}

\end{document}